\makeatletter
\declare@file@substitution{revtex4-1.cls}{revtex4-2.cls}
\makeatother
\documentclass[twocolumn,twocolappendix]{aastex631}
\usepackage{amsmath}

\shorttitle{Cycle Phase Dependence of Magnetic Power Spectra}
\shortauthors{Yukun Luo et al.}
\graphicspath{{./}{figures/}}
\usepackage{threeparttable}
\usepackage{booktabs}

\begin{document}

\title{The Sun's Magnetic Power Spectra over Two Solar Cycles. \uppercase\expandafter{\romannumeral2}. Cycle Dependence of Active Region, Magnetic Network, and Their Relation}

\correspondingauthor{Jie Jiang}
\email{jiejiang@buaa.edu.cn}

\author{Yukun Luo}
\affiliation{School of Space and Environment, Beihang University, Beijing, People’s Republic of China}
\affiliation{Key Laboratory of Space Environment Monitoring and Information Processing of MIIT, Beijing, People’s Republic of China}

\author{Jie Jiang}
\affiliation{School of Space and Environment, Beihang University, Beijing, People’s Republic of China}
\affiliation{Key Laboratory of Space Environment Monitoring and Information Processing of MIIT, Beijing, People’s Republic of China}

\author{Ruihui Wang}
\affiliation{School of Space and Environment, Beihang University, Beijing, People’s Republic of China}
\affiliation{Key Laboratory of Space Environment Monitoring and Information Processing of MIIT, Beijing, People’s Republic of China}

\begin{abstract}

The multi-scaled solar magnetic field consists of two major components: active regions (ARs) and magnetic network. Unraveling the cycle-dependent properties and interrelations of these components is crucial for understanding the evolution of the solar magnetic field. In this study, we investigate these components using magnetic power spectra derived from high-resolution and continuous synoptic magnetograms since cycle 23 onwards. Our results show that the size of the magnetic network ranges from 26 Mm to 41 Mm without dependence on the solar cycle. The power of the network field ($P_{NW}$) accounts for approximately 20\% of the total power during any phase of solar cycles. In contrast to the AR power ($P_{AR}$), $P_{NW}$ displays a weaker cycle dependence, as described by the relationship $P_{NW}$ $\approx$ 0.6* $P_{AR}$ + 40. The power-law index between AR sizes and magnetic network sizes presents a strong anti-correlation with the activity level. Additionally, our study indicates that in the absence of sunspots on the solar disc, the magnetic power spectra remain time-independent, consistently exhibiting similarity in both shape and power. This study introduces a new method to investigate the properties of the magnetic network and provides magnetic power spectra for high-resolution simulations of the solar magnetic field at the surface at various phases of solar cycles.
\end{abstract}


\section{Introduction} \label{sec:intro}

Solar magnetic fields, as the main source of solar activity, include two prominent components: active regions (ARs) and magnetic network \citep{Solanki2006}. As typical magnetic structures on the solar surface, they also interact with various scale flow fields like differential rotation and supergranulation flows \citep{Rincon2018}. The evolution of ARs may also influence the distribution and formation of network fields. The investigation of their interaction and temporal variation with the solar cycle can help to understand the evolution of magnetic fields, dynamic processes with flow fields, and their impact on the solar atmosphere \citep{Rubio2019, Lidia2015}.

As the most obvious magnetic configurations on the solar surface, ARs exhibit substantial scales and strong enough magnetic fields to yield pronounced, easily detectable signals \citep{Lidia2015}. Decades of continuous observations of the solar magnetic fields have resulted in a wealth of data on ARs. These lead to the well-studied cycle dependence of ARs. However, we know little about magnetic networks due to their small spatial scales and relatively rapid evolution \citep{Rubio2019}. Consequently, there is an ongoing debate about the properties of the magnetic network.

The first ongoing debated question is the origin of the magnetic network. According to \cite{Martin1990}, ephemeral regions (ERs) contribute 90$\%$ or more of the network flux, with the remaining flux originating from the internetwork fields. \cite{Schrijver1997} build upon this conjecture and construct a model to investigate the evolution of the network flux, ignoring the contribution of the magnetic internetwork. Their findings support ERs as the main source of the network, with only a small amount of flux dispersing from ARs. \cite{Hagenaar2003} conduct further investigation and suggest the dominance of ERs in the relatively small network, while the larger scale network results from decaying ARs. However, \cite{Gosic2014} argue that the internetwork is the most important source, not ERs. As these efforts present different potential sources of the network flux, further investigation is needed.

The second open question is whether the network flux and size vary with the solar cycle. Answering this question is not only necessary to understand the formation and sustainment of the network but also to shed light on the process of flux diffusivity from large concentrations to the network \citep{Leightonn1964}. \cite{Hagenaar2003} suggest that the network with fluxes $\le2\times10^{19}$ Mx is independent of solar cycles, while the stronger network increases with solar activities because of the antiphase relationship between ERs and ARs. However, we still know little about the variation of total network flux \citep{Meunier2003}. 

Previous attempts to determine the variation of network size with the solar cycle use different proxies, with divergent results. \cite{Hagenaar1997}, focusing on the chromospheric network, propose no dependence of network sizes on local magnetic strength. Both regarding the chromospheric network, \cite{Singh1981, Berrilli1999} find a tendency toward smaller network sizes at the solar maximum, while \cite{Wang1988,Muenzer1989} report a contrary trend. Focusing on supergranulation, \cite{Derosa2004} conduct studies direct from Dopplergram and support the anti-correlated cycle dependence of supergranulation sizes. \cite{Meunier2007} use intensity maps to study the variation of the supergranulation size with different parts of magnetic fields. They find that larger supergranulation is associated with stronger network fields, but sizes decrease with increasing magnetic strength within supergranulation. In addition to observational approaches, simulation-based studies by \cite{Crouch2007,Thibault2012,Thibault2014} employ the diffusion-limited aggregation model to support the positive dependence of characteristic network sizes on solar activity. 

However, proxies on magnetic network studies may introduce unexpected discrepancies. For example, the dynamical interaction between supergranulation and magnetic network is not well understood \citep{Rincon2018}. The magnetic network has a large uncertainty range in relation to the chromospheric network \citep{Schrijver1996}. Investigating the magnetic network based on magnetograms can directly answer whether the network properties vary with the solar cycle. The power spectrum obtained from magnetograms is a useful tool to identify and study magnetic structures and can be used to measure the typical scale at which the magnetic fields are organized. Magnetic power spectra obtained by \cite{Abramenko2001} reveal suspected network structures. In the first paper of the series, \cite{Luo2023} (hereafter referred to as Paper \uppercase\expandafter{\romannumeral1}) also identify magnetic structures at supergranulation scales within the power spectra. Hence, we can expect to identify the network in magnetic power spectra and investigate network properties and origin based on the power spectra features related to the network, such as scale and power-law index.

As the second paper in the series, we improve our identification methods for determining AR sizes and network sizes from magnetic power spectra. The analysis is extended to cover the entire solar cycles 23, 24, and part of cycle 25. We use Solar Dynamics Observatory (SDO)/Helioseismic and Magnetic Imager (HMI) and Solar and Heliospheric Observatory (SOHO)/Michelson Doppler Imager (MDI) synoptic magnetograms. The calibration method proposed by Paper \uppercase\expandafter{\romannumeral1} is applied to ensure comparable and homogeneous identification results. Our goal is to examine how the size and power of the network vary with solar cycles, the relation between ARs and the network, and to attempt to reveal the possible origin of the magnetic network. Additionally, the spotless days during the three solar minimums provide an opportunity to speculate on the network properties during grand minima.

This paper is organized as follows. In Section \ref{sec:datas}, we describe the HMI and MDI synoptic magnetograms and improve the methods for determining AR sizes and network sizes. Section \ref{sec:results} subsequently presents the identification results and explains variations of the magnetic network with solar cycles. The relation between AR power and network power is also presented in this section. We summarize and discuss our results in Section \ref{sec:conclusion}. 

\section{Data and Methods}\label{sec:datas}
\subsection{Data} \label{subsec:data}

This paper uses radial synoptic magnetograms observed by MDI on board SOHO \citep{Scherrer1995} and HMI on board SDO \citep{Scherrer2012, Scherrer2012b}, respectively. They cover cycles 23, 24, and part of cycle 25, beginning in Carrington Rotation (CR) 1911 (1996 July) and ending in CR 2265 (2022 December). The data sets include a total of 355 CRs.

Same as Paper \uppercase\expandafter{\romannumeral1}, we utilize the `pyshtools' package in Python to perform spherical harmonic decomposition for synoptic maps \citep{Wieczorek2018,Wieczorek2019}, subsequently deriving their magnetic power spectra. To meet the requirements of the algorithm in the package, the grid size must be $n \times n$ or $n \times 2n$. Therefore, we transform the data resolution of MDI from 1080$\times$3600 to 1080$\times$2160, resulting in the maximum spherical harmonic degree $l_{max}$ is 539. Similarly, for HMI data, the resolution is transformed from 1440$\times$3600 to 1440$\times$2880 and $l_{max}=719$. To ensure a consistent analysis, we calibrate all HMI magnetic power spectra by the method proposed in Paper \uppercase\expandafter{\romannumeral1}. Our analysis focuses only on MDI and HMI power spectra between the range $l=6$ and 539.

\subsection{Determining AR Sizes and Network Sizes} \label{subsec:method}

\begin{figure*}[ht]
    \centering
    \includegraphics[width=1.0\linewidth]{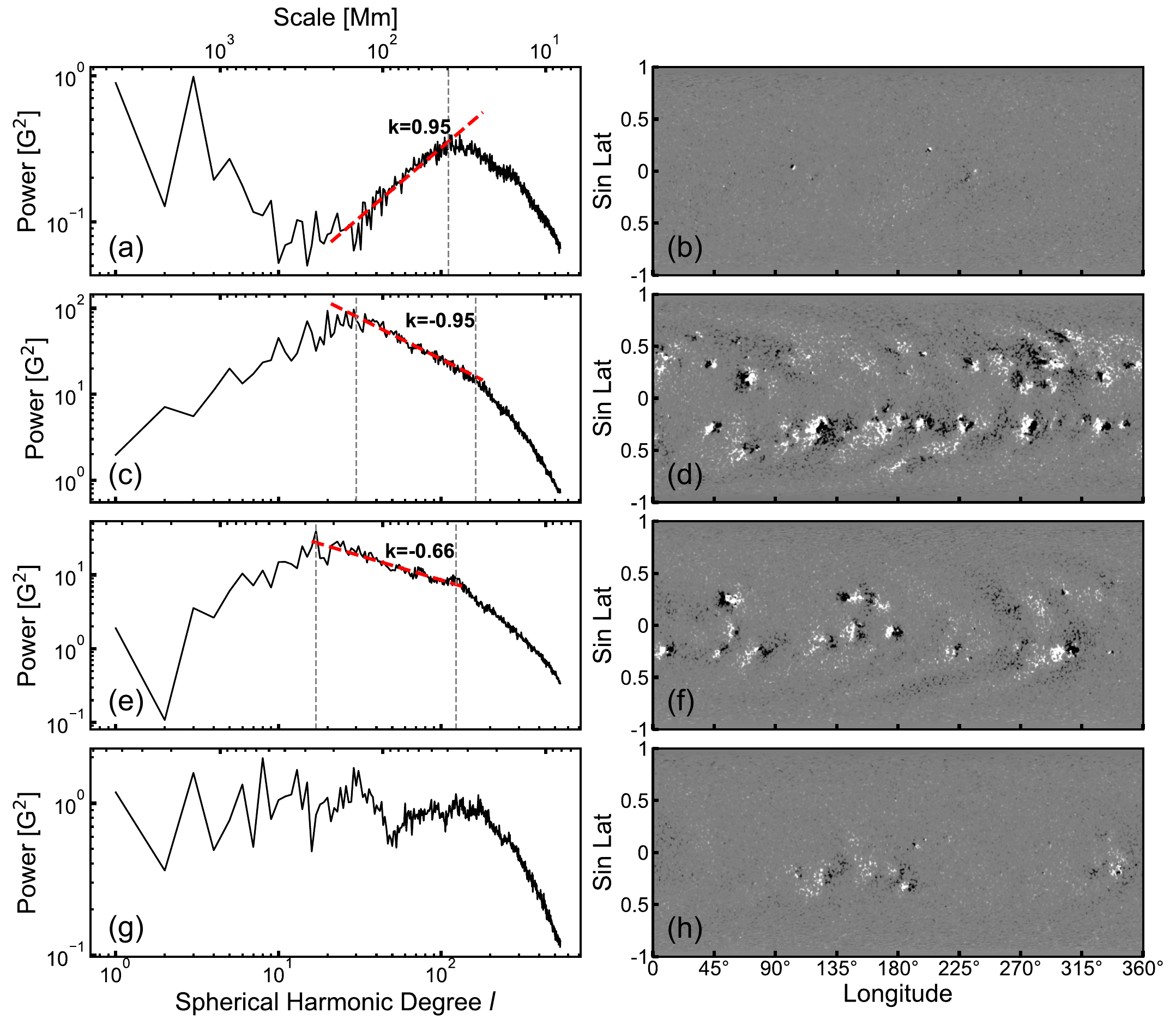}
    \caption{Examples of the four types of magnetic power spectra (left) and the corresponding synoptic maps (right). From top to bottom are CRs 2063, 1960, 2014, and 2048, respectively. The saturation values of synoptic maps are all 150 G.} Identified AR sizes and network sizes are marked by the vertical gray dashed line. The red dashed lines are the fitting ones of power spectra between AR sizes and network sizes.
\label{fig:fig1}
\end{figure*}

Based on the relative strength between AR power and network power, there are four types of magnetic power spectra. Figure \ref{fig:fig1} shows the typical examples for each type. The typical supergranular size is generally regarded as approximately 36 Mm \citep{Williams2014, Hathaway2015, Rincon2018}, so we restrict the identification range to be $l=105\sim170$ (26 Mm $\sim$ 42 Mm). This aims to minimize the effects caused by similar scale structures like ERs. We'll discuss the effect of different identification ranges in Section \ref{subsubsec:synoptic}. Below are the four types of magnetic power spectra, based on which we identify the typical network size using different methods.

The top panel of Figure \ref{fig:fig1} shows the first type. During the solar minimum, there are no AR features, only a peak corresponds to the network. We identify the positions of the peaks through an available peak-finding algorithm: `scipy.signal.find$\_$peaks' within Python \citep{virtanen2020scipy}. If there is only one local maximum that meets the prominence and width threshold within the identification range, it is the peak we look for. 
Using this algorithm, we identify network sizes for 34 CRs.

During the active phase of the solar cycle, the AR power is much stronger than the network power. As a result, the network appears as the knee, as shown in the second line of Figure \ref{fig:fig1}. For this type, we use an algorithm similar to \cite{satopaa2011} to identify the location of the network. This algorithm detects the knee by finding the point with the most significant change in slope, based on the Lagrange median theorem. To do this, one endpoint of the identification range is fixed, and as the other endpoint moves, the knee is identified as the point that is parallel to the secant line and appears to be unmoving. In this type, network sizes for 103 CRs are identified.

If the AR power slightly exceeds the network power, the magnetic network signal will appear as a small peak instead of a knee. The third line of Figure \ref{fig:fig1} is an example of this third type. To enhance the peak signal, which is easily affected by spectral fluctuations, we subtract the baseline of the power spectrum. We use the asymmetric least squares method in the `pybaselines' package to obtain the baseline \citep{Erb2024}. We identify the peak in the processed power spectrum using two morphological methods. The first method is the same as the one used for the first type. The second method is developed by \cite{Duarte2021}. Their method selects peaks based on multiple parameters such as the distance and height difference between neighboring peaks. The peak remains only if it is identified by both methods. The third type has 117 CRs with identified network sizes. 

Some power spectra may not exhibit significant characteristic sizes, or there may be multiple signals strong enough to be identified within the given range. For example, in the bottom panel of Figure \ref{fig:fig1}, the AR power has the same intensity as the network power. There are multiple peaks that may correspond to ERs or fragments of ARs. We can not distinguish which is the magnetic network so all peaks are not identified as magnetic network. To ensure the accuracy of our methods, we exclude this type, i.e., the fourth type of power spectra from the identification process, resulting in the exclusion of 101 CRs.

For ARs, we restrict the identification range for $l=10\sim60$ (73 Mm $\sim$ 438 Mm) to reduce possible misidentifications. During the active phase, the power spectrum shows the AR power as the strongest signal, appearing as a peak (see Figures \ref{fig:fig1} (c) and (e) for examples). We use the same algorithm as for the first type of network identification. The locations of identified ARs are marked by the left dashed vertical lines in the plots. We have identified ARs in the power spectra of 196 CRs. If the AR is too weak like Figure \ref{fig:fig1} (a), or if there are interference signals as shown in Figure \ref{fig:fig1} (g), we exclude the power spectrum from the identification process. 

\section{results} \label{sec:results}
\subsection{AR Typical Size Detected Based on Power Spectrum}\label{subsec:AR}
\begin{figure}[ht]
    \centering
    \includegraphics[width=1.0\linewidth]{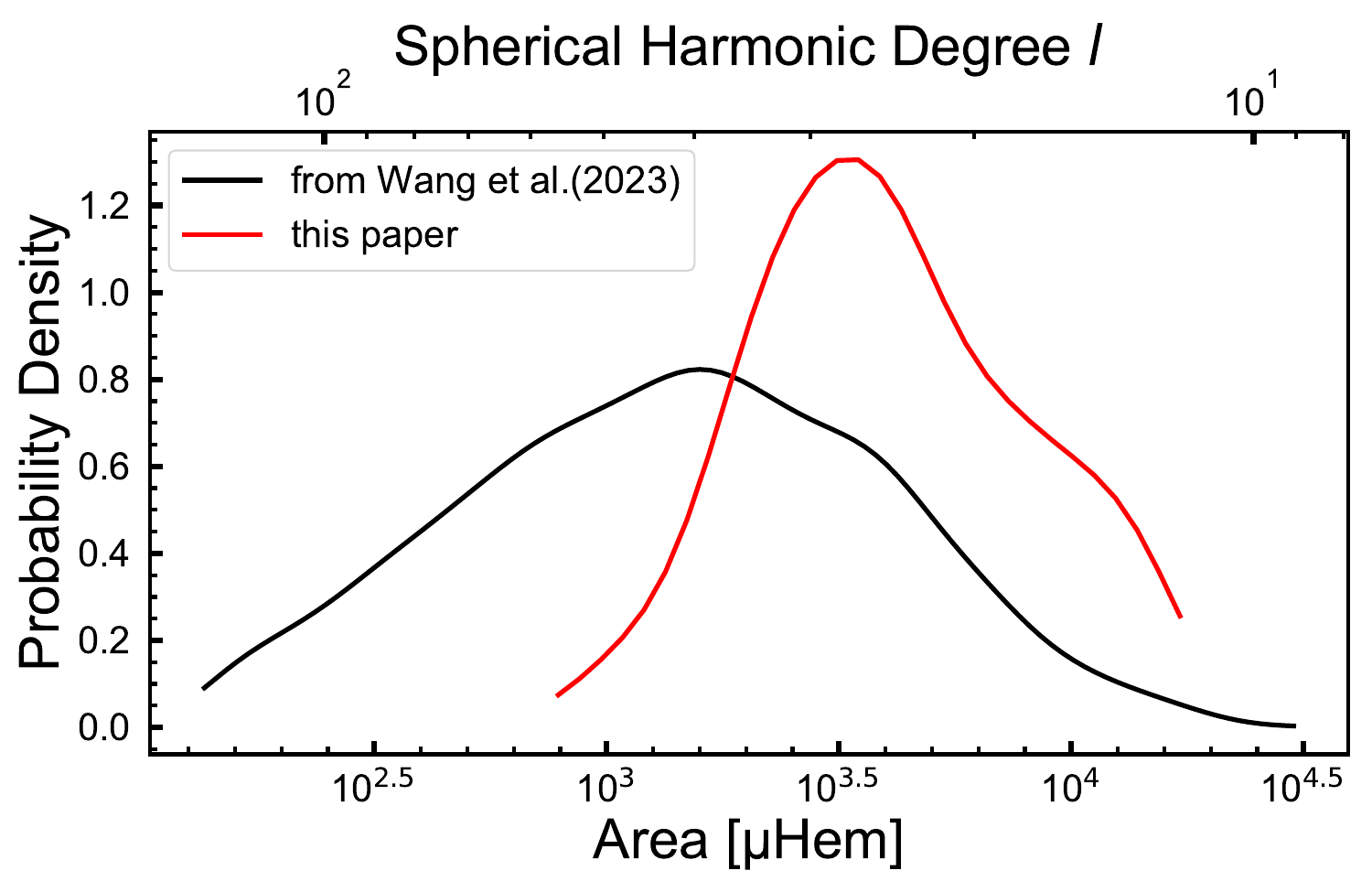}
    \caption{Probability density functions of AR areas obtained by the database of \cite{Wang2023} (black) and power spectra (red), respectively.}
    \label{fig:fig2}
\end{figure}

The size of ARs we identified ranges from $l=12$ (365 Mm) to $l=56$ (78 Mm). Assuming that the identified ARs are both the circular bipolar magnetic region (BMR), we can convert the size information into areas using the following equation:
\begin{equation}\label{eq1}
     Area = 2\pi {(\frac{1}{4}\frac{{2\pi R_{\odot}}}{l})^2}, 
\end{equation}
where $R_{\odot}$ is the radius of the Sun and $l$ is the spherical harmonic degree. The area of ARs in our results ranges from 786 $\mu Hem$ to 17135 $\mu Hem$. The mean area is 5135 $\mu Hem$ and the corresponding size is about 190 Mm ($l\approx22$). The probability density function (PDF) of areas is shown as the red line in Figure \ref{fig:fig2}. The peak of the PDF is located at 3481 $\mu Hem$, corresponding to 165 Mm ($l\approx27$), which is slightly smaller than the mean value.

In previous work, ARs are usually detected morphologically from magnetograms and then their sizes are measured. Our recent AR database from \cite{Wang2023} is a typical example, which is used here for comparison. The corresponding PDF is shown as the black line in Figure \ref{fig:fig2}. Both PDFs are close to the log-normal distribution commonly found in previous work about ARs \citep{Abramenko2005,Canfield2007}. However, the PDF obtained in this paper has a narrower full width at half maxima, indicating that ARs identified from power spectra tend to concentrate on a narrow range. Additionally, the mean area and the peak of the PDF from the database are 2360 $\mu Hem$ ($\approx$135 Mm, $l\approx32$) and 1584 $\mu Hem$ ($\approx$110Mm, $l\approx39$), respectively, which are significantly lower than the values of the red line. 

Two aspects may contribute to the discrepancy in mean areas. ARs have irregular shapes. The AR scale identified in magnetograms morphologically corresponds to the total number of pixels of the irregular structure. In contrast, the scale identified in a power spectrum is indicative of the maximum distance between the edges of a network. Therefore, the areas obtained by Equation (\ref{eq1}) are systematically larger than those obtained by morphological methods. The second reason is that the stronger signal in the power spectrum will mask the weaker one. Based on the positive relationship between AR areas and flux proposed by \cite{Sheeley1966}, the smaller ARs are usually related to weaker ones. Hence, the relatively small ARs tend to be masked, resulting in systematically larger identified sizes.

\subsection{Cycle Phase Dependence of Typical Network Size}\label{subsec:SG}
\subsubsection{Identification Based on Individual Synoptic Maps}\label{subsubsec:synoptic}
\begin{figure}[ht]
    \centering
    \includegraphics[width=1.0\linewidth]{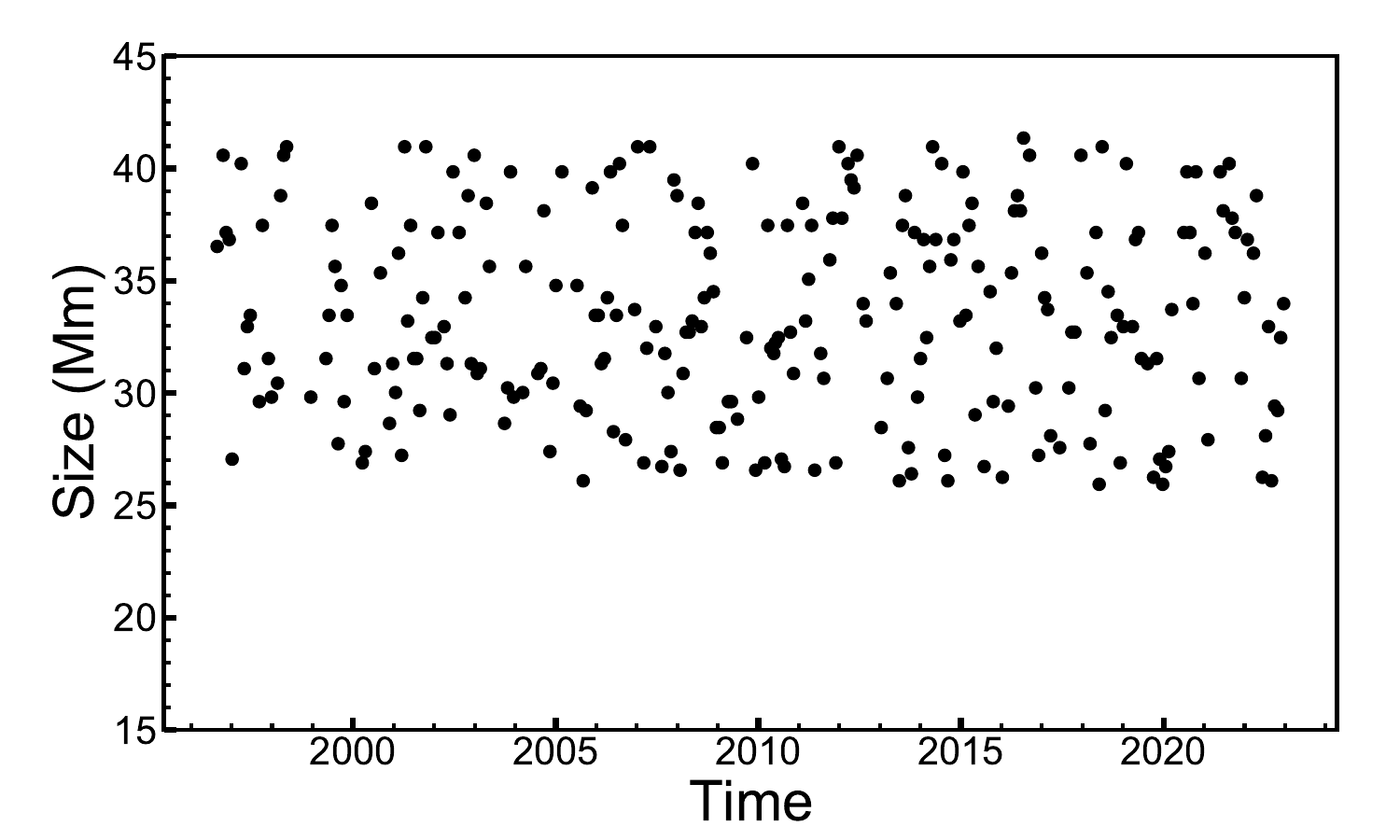}
    \caption{Network sizes identified in magnetic power spectra during solar cycles 23, 24, and part of 25.}
    \label{fig:fig3}
\end{figure}

\begin{figure}[ht]
    \centering
    \includegraphics[width=1.0\linewidth]{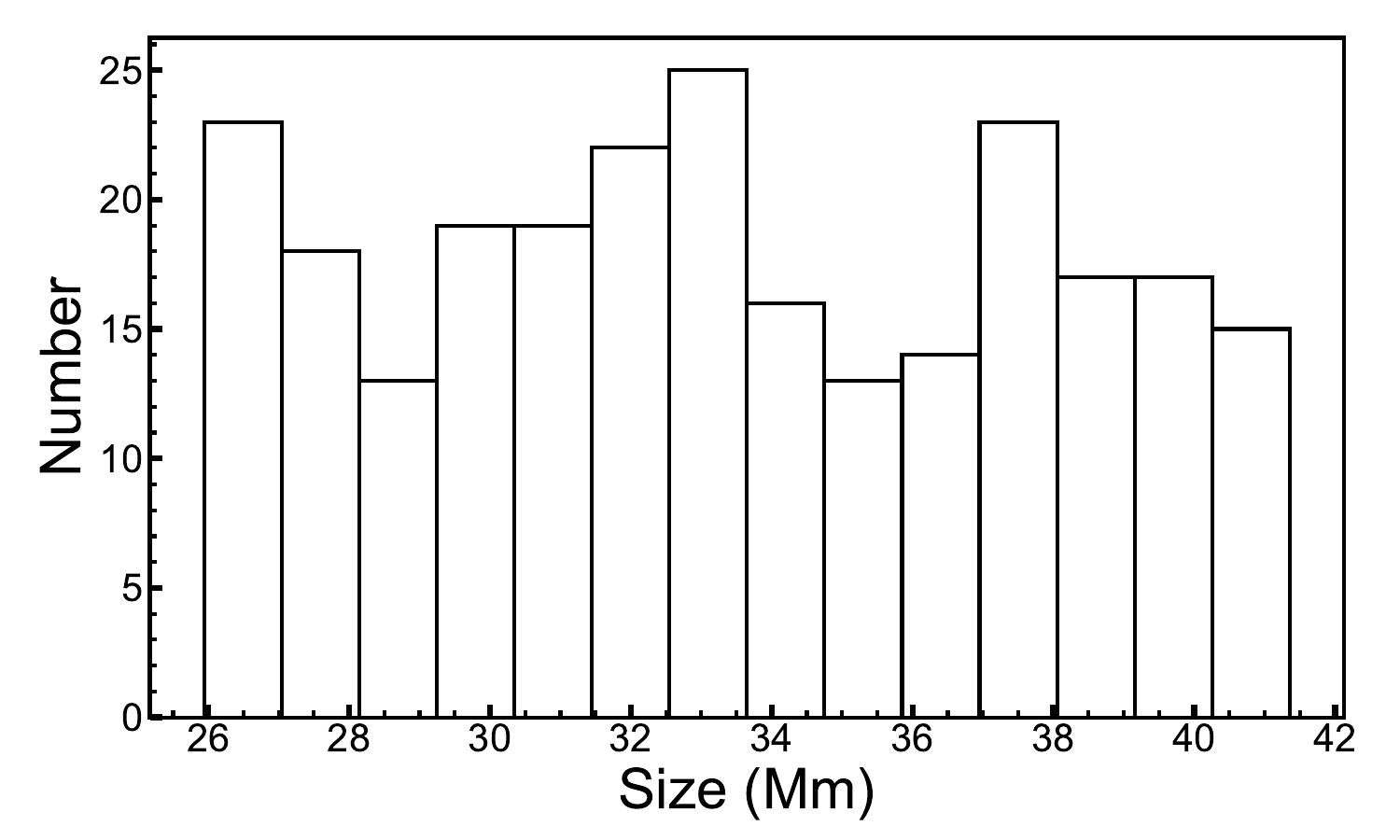}
    \caption{Histogram of network sizes identified in magnetic power spectra.}
    \label{fig:fig4}
\end{figure}

We first investigate the cycle phase dependence of typical network size based on individual synoptic maps. Figures \ref{fig:fig1} (a), (c), and (e) show 3 typical examples. The vertical dashed lines on the right highlight the identified network. The identified network sizes range from $l=106$ ($\approx$26 Mm) to $l=169$ ($\approx$41 Mm) and their time dependence is shown in Figure \ref{fig:fig3}. The scale range is broad and nearly homogeneous in any phase of the solar cycle, implying that the probability of the network emerging at any size within the scale range is approximately equal. Meanwhile, this emphasizes that the network is also obvious in the active phase not only during solar minimum. Additionally, there is no significant variation in the scale range for the period. This is an indication that the network sizes have no significant cycle phase dependence or cycle dependence. It is worth noting that there are some spotless days during three solar minimums, implying that the size of the magnetic network may also have similar properties during the solar grand minimum.

Although the results presented above come from two datasets with different spatial resolutions and magnetogram sensitivities, the analysis in Paper \uppercase\expandafter{\romannumeral1} shows that the discrepancy in the spherical harmonic degree of the network size between the HMI and MDI synoptic maps, $\Delta l$, is less than 5 after the calibration. The discrepancy between the two datasets has a negligible impact on the identification of AR and network sizes.

Figure \ref{fig:fig4} shows the histogram of identified network sizes. The mean network size is 33.41 Mm ($l=131$), with a standard deviation of 4.43 Mm. The skewness and kurtosis are 0.058 and 1.87, respectively. These values are similar to the characteristic parameters of a uniform distribution. This suggests that our result is close to a uniform distribution within the identification range. However, previous work on supergranulation typically shows a unimodal distribution, as shown by \cite{Simon1964} using the autocorrelation method. This may be contributed by the same reason mentioned in Section \ref{subsec:AR}: the distribution we obtain pertains to the typical network size, which contains the strongest network power. In addition to this reason, there are some slight differences between supergranulation and magnetic network \citep{Rincon2018, Requerey2015, Requerey2017}. \cite{Hagenaar1997} report that the sizes estimated from magnetograms are typically smaller than Dopplergrams. Therefore, the network size and the supergranulation size could present different distributions.

The effects of the various identification ranges of $l$ on the typical network size and its cycle phase dependence, as mentioned in the beginning of Section \ref{subsec:method}, are presented in the Appendix. If using a broader identification range like $l=100\sim300$, the identified results smaller than 25 Mm mainly concentrate on 15 Mm. ERs might contribute to the high concentration of 15 Mm because the typical ER size is smaller than about 22 Mm \citep{Tang1984,Harvey1993}. This result indicates that the upper limit $l$ of the identification range should be set to the size corresponding to approximately 25 Mm. When selecting the upper limit of the identification range from $l=170$ (24.5 Mm) to $l=180$ (25.9 Mm), the identified network sizes exhibit minimal variation and remain cycle-independent. To minimize the potential effects of ERs, we choose 24.5 Mm, i.e., $l=170$ as the upper limit. Similarly, the lower limits of the identification range of $l=100, 105$, i.e., 43.8 Mm and 41.75 Mm, yield comparable outcomes.

Some structures of a similar scale like ERs can introduce random interference in power spectra, bringing the uncertain range of our identification results. A weak cycle dependency of network sizes could be masked by the uncertainty. This bias can be reduced by smoothing and averaging power spectra, which will be investigated in the next subsection. 

\subsubsection{Identification Based on Averaged Spectra over Maximum and Minimum Phases}\label{subsubsec:average}

\begin{figure}[ht]
    \centering
    \includegraphics[width=1.0\linewidth]{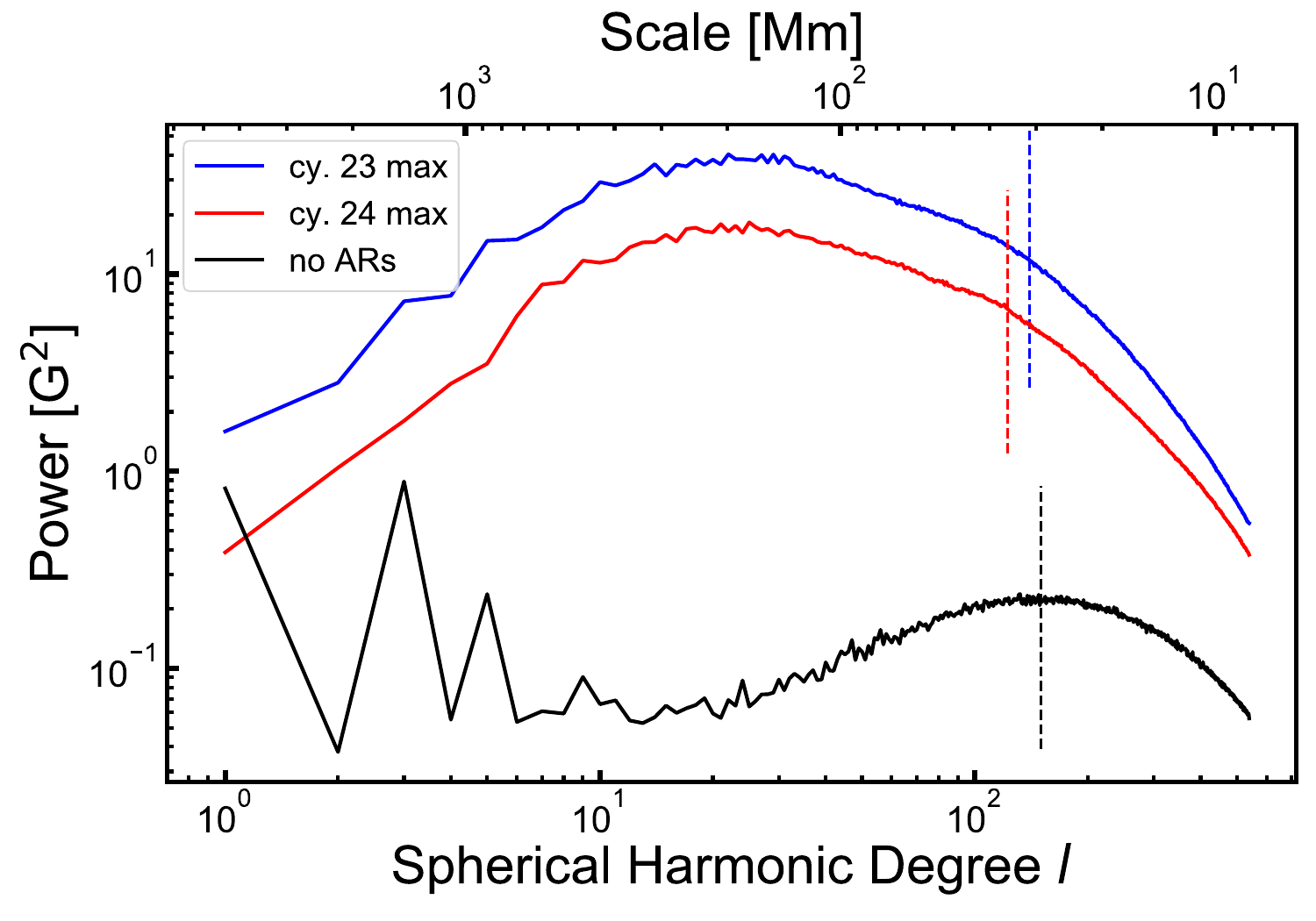}
    \caption{Averaged magnetic power spectra at different phases of solar cycles. Blue: cycle 23 maximum; red: cycle 24 maximum; black: solar minimum across three solar cycles without any ARs. The vertical dashed lines mark the network sizes identified.}
    \label{fig:fig5}
\end{figure}

The identified results in average power spectra during different phases are displayed in Figure \ref{fig:fig5}. Data in the active phase are selected using the monthly smoothed total sunspot number (SSN, version 2.0) as a threshold. We choose $SSN > 100$ for cycle 23 (blue lines) and $SSN > 64.5$ for cycle 24 (red lines). The typical network power is 11.8 $G^2$ and 6.7 $G^2$, respectively. For the quiet phase (black lines), we use synoptic maps without any spots as the representation. To account for the limited data that satisfies the conditions, we combine data from various solar minimums: CRs 2073, 2074, 2082, 2210, 2222, 2223, and 2227. The potential difference of power spectra between various solar minima is negligible, which is discussed in Section \ref{subsubsec:minimum}. The typical network power only reaches 0.2 $G^2$, which is significantly lower than that during the solar maximum. Section \ref{subsec:dependence} will examine the meaning and reason for this difference in more detail.

The typical network sizes for cycle 23 maximum, cycle 24 maximum, and the solar minimum are 31.3 Mm, 35.9 Mm, and 29.2 Mm, respectively. However, this data is not sufficient to support the idea that network size increases with solar activity. It should be noted that the peak in the average power spectrum of the solar minimum is relatively flat. This means that the power of magnetic structures near the peak is also strong enough to be considered as network features. Hence, we believe that the range of typical network sizes can be extended and even to include the size during the solar maximum.

Combining the above analysis, the conclusion of Section \ref{subsubsec:synoptic} can be further confirmed: the network size is not dependent on the solar cycle. A similar result is given by \cite{Hagenaar1997}, who focused on the chromospheric network and also suggested no dependence of the network size on the local magnetic strength. Next, we will discuss the cycle dependence of the network power and its relation with ARs.

\subsection{Cycle Phase Dependence of the Power Index for the Range between AR Size and Network Size}\label{subsec:slope}

\begin{figure}[ht]
    \centering
    \includegraphics[width=1.0\linewidth]{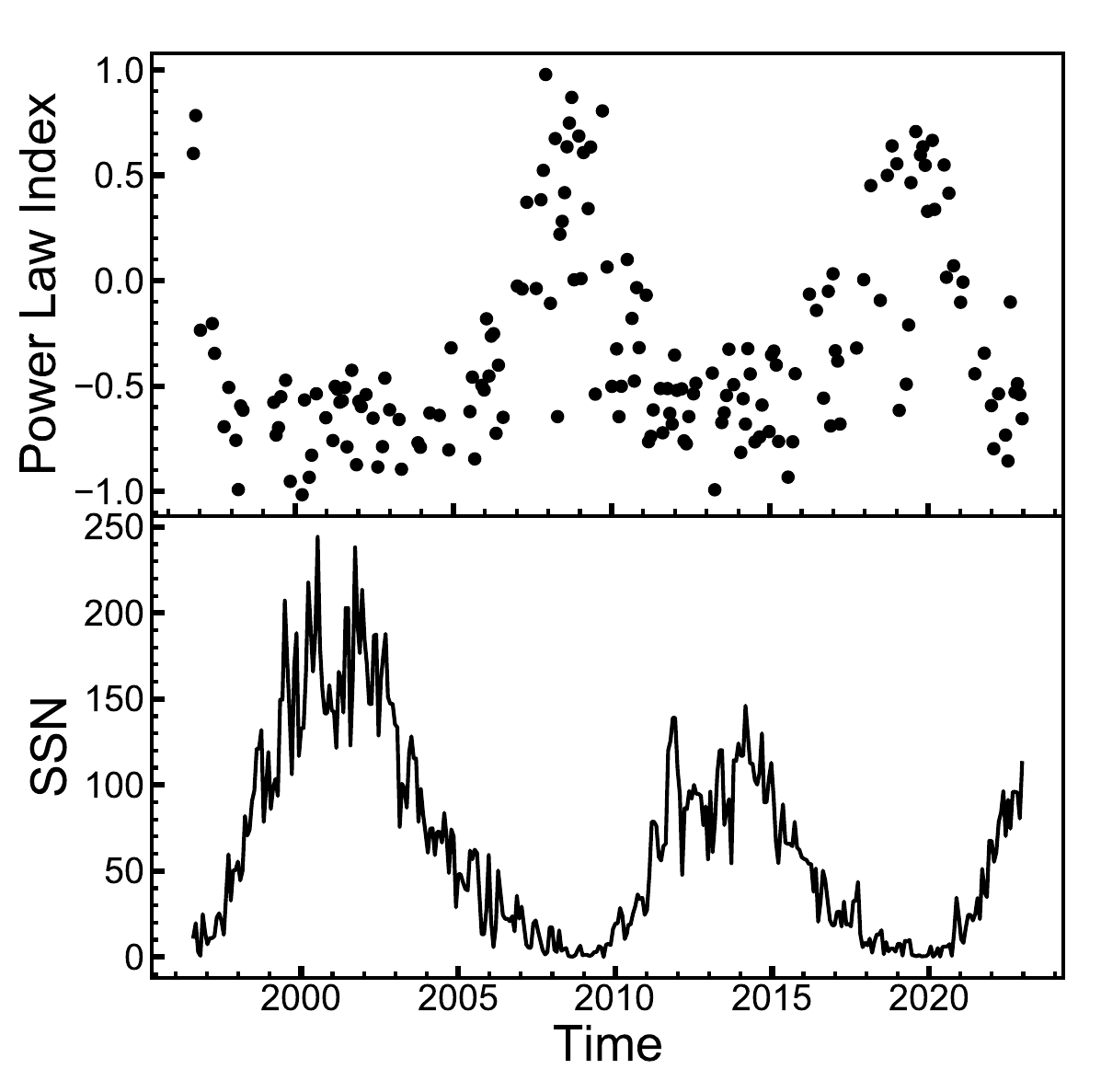}
    \caption{Time evolution of power-law indices (top) between AR sizes and network sizes and SSN (bottom).}
    \label{fig:fig6}
\end{figure}

\begin{figure}[ht]
    \centering
    \includegraphics[width=1.0\linewidth]{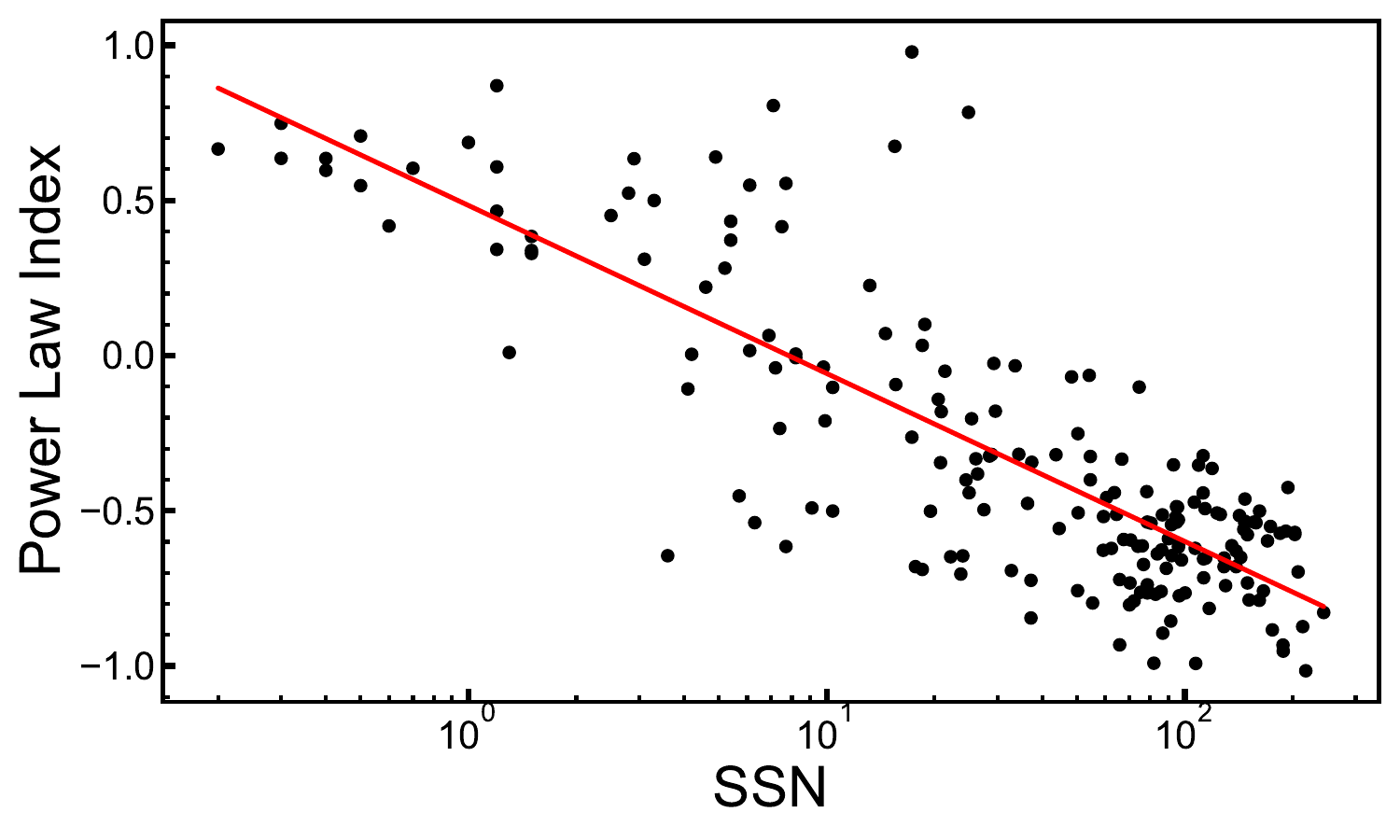}
    \caption{Power-law indices versus SSN. The red line is the fitting line.}
    \label{fig:fig7}
\end{figure}

In Figures \ref{fig:fig1} (c) and (e), the power spectra between AR sizes and network sizes are nearly linear. The linear power spectra suggest that the cascade from AR power to network power is universal. The AR could be one of the sources of the network, and the diffusivity could be homogeneous across multiple scales. The power-law index between AR sizes and network sizes can help investigate their relation.

To determine the power-law index through the least square fitting, we use the sizes of ARs and the network from the previous subsections as range endpoints. During the solar minimum, ARs are rare, so we use $l=30$ as the left endpoint of the fitting range for this period. To avoid biases caused by the shape of peaks, we choose $l=l_{NW}-10$ as the right endpoint, where $l_{NW}$ is the location of the identified network. We judge the goodness of fit using the residual sum of squares (RSS) as a criterion. We exclude power-law indices with $RSS > 0.55$, leaving 191 CRs with indices ranging from -1.02 to 0.98, as shown in Figure \ref{fig:fig6}. The comparison between the top panel and the bottom panel shows a significant anti-correlation between the indices and the solar cycle. Figure \ref{fig:fig7} displays the scatter plot between indices and SSN. The power-law indices have a linear relationship with the logarithm of SSN. Through least square fit, we obtain the following formulation: 
\begin{equation}\label{eq2}
\begin{array}{c}
k=(-0.54\pm0.03)*\lg{SSN}+0.48\pm0.05, 
\end{array}
\end{equation}
where $k$ is the power-law index. This equation provides a way to evaluate the power-law index with a given SSN.

Equation (\ref{eq2}) shows that the solar activity modulates the power-law index: stronger activity results in a smaller index. During the solar active phase of cycle 23, the power-law indices are universally smaller than those in cycle 24. The average indices for solar active phases are -0.65 and -0.59, respectively. This is due to cycle 23 being stronger than cycle 24. However, all indices are smaller than -1.5, as proposed by \cite{Kraichnan1965} about magnetohydrodynamics turbulence power spectra. This suggests the presence of energy injection in network scales.

As mentioned in the Introduction, network features are present in the magnetic power spectra of \cite{Abramenko2001}. The power-law index they obtained also corresponds to the range between AR sizes and network sizes (see Figure 7 in their paper). For local AR magnetograms, their indices are $-0.71\pm0.02$, which is close to the indices shown by the red dashed lines in Figures \ref{fig:fig1} (c) and (e). The results for local magnetograms of quiet sun are $0.32$ and $0.49$, smaller than the value of $0.95$ shown in Figure \ref{fig:fig1} (a), but within the range of variation we obtained. This supports our identification and fitting results.

According to Paper \uppercase\expandafter{\romannumeral1}, the difference in the quality of HMI and MDI magnetograms has a negligible impact on the power law indices after calibration. The variation of power-law indices is controlled by both ARs and magnetic network. In the next section, we will quantitatively examine how these two structures affect the indices.

\subsection{Understanding the Cycle Phase Dependence of the Power Law Index}\label{subsec:dependence}
\subsubsection{Cycle Phase Dependence of Network and AR Powers}\label{subsubsec:dependency}

\begin{figure}[ht]
    \centering
    \includegraphics[width=1.0\linewidth]{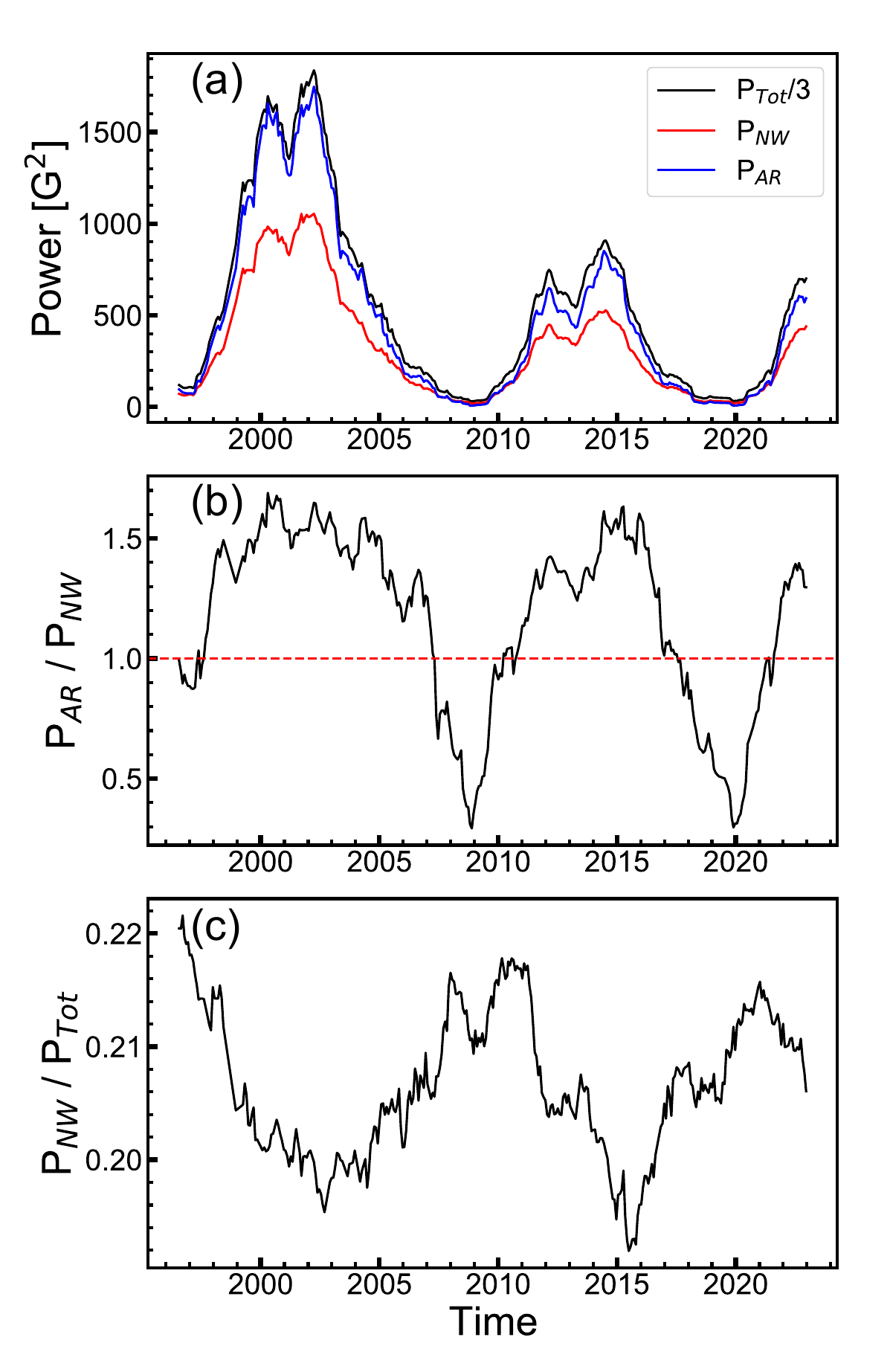}
    \caption{Comparison of different parts magnetic power. (a): the evolution of total power (black), network field power (red), and AR power (blue). (b): the variation of ratios between AR power and network power with the solar cycle. (c): similar to (b) but for the ratio between network power and total power. The data are smoothed using a smoothing window of two years.}
    \label{fig:fig8}
\end{figure}

\begin{figure}[ht]
    \centering
    \includegraphics[width=1.0\linewidth]{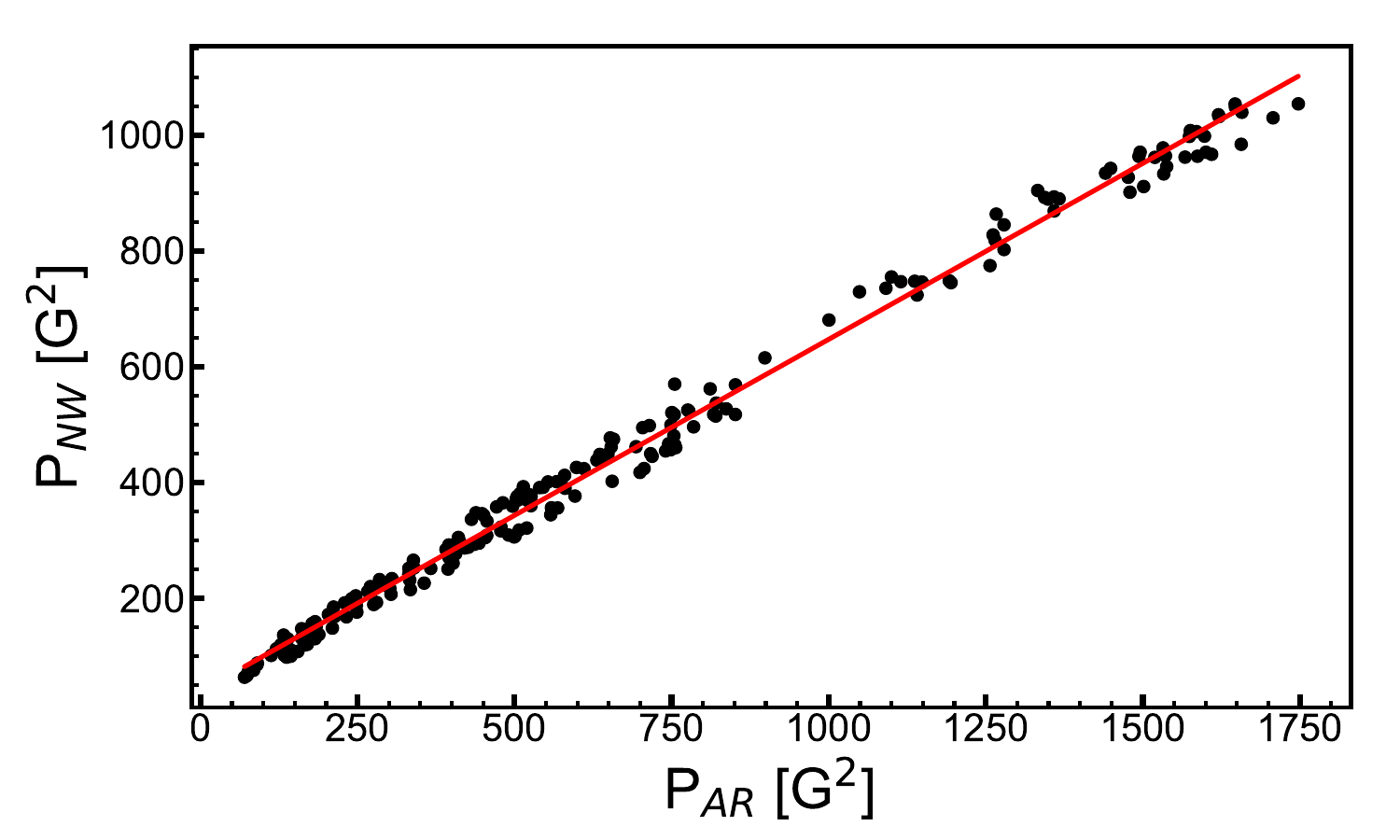}
    \caption{Magnetic network power versus AR power during the solar active phase.}
    \label{fig:fig9}
\end{figure}

We use ${P_{AR}} = \sum\limits_{l = 10}^{50} {{P_l}}$ as the AR power and ${P_{NW}} = \sum\limits_{l = 105}^{180} {{P_l}}$ as network power, where $P_l$ is the power for spherical harmonic degree $l$. The time evolution of these powers, as well as that of the total power ${P_{Tot}} = \sum\limits_{l = 1}^{539} {{P_l}}$, is displayed in Figure \ref{fig:fig8} (a). The total power is divided by three for comparison. In general, the AR and network power vary with the total power, with the former being stronger than the latter most of the time. Only during the solar minimum is the network power stronger. The minimum value of network power is 19.5 $G^2$, which occurs during the solar minimum between cycle 24 and cycle 25. The maximum value is 1054 $G^2$. It has increased by 54 times. This amplitude is larger than expected from previous work \citep{Hagenaar2003}. The AR power has a minimum value of 6 $G^2$ and a maximum value of 1748 $G^2$. The increasing factor can be as high as 288, significantly larger than that of network power.

In Figure \ref{fig:fig8} (b), a comparison is made between AR power and network power. The ratio has a significant solar cycle dependency, with the AR power being only 0.3 times the network power during the solar minimum. As the ARs emerge frequently, the ratio increases rapidly until it reaches 1.6 during both the solar maximum for cycle 23 and cycle 24. The ratio is dominated by AR power and gradually saturates when it approaches 1.6.

To investigate the relationship between decaying ARs and magnetic networks, we analyze data from the active phase above the red line in Figure \ref{fig:fig8} (b). The comparison between AR power and network power during this phase is shown in Figure \ref{fig:fig9}. The relationship is 
\begin{equation}\label{eq3}
    P_{NW}=(0.61\pm0.003)*P_{AR}+(39.66\pm2.86). 
\end{equation}
The intercept term may represent the sources for network flux that are independent of solar cycles. We suggest these sources could be ERs or internetwork fields, which are thought to have little or no variation with solar cycles. The power from these sources is only about 40 $G^2$, significantly smaller than the network power during the solar maximum. This might imply that the decaying AR is the dominant contributor to network fluxes during the solar maximum. Figure \ref{fig:fig8} (b) shows that during the minima of solar cycles 23 and 24, the magnetic power in ARs is just about one third of the power in networks. Previous studies \citep[e.g.,][]{Harvey1975,Schrijver1994,Chae2001,Hagenaar2001,Jin2015} indicate that internetwork fields and ERs bring a huge amount of flux to the solar surface. Internetwork fields and ERs could also be source of the network field during the solar minima. 

The fixed linear relationship may indicate that cancellation between flux concentrations dissipates approximately $40\%$ of AR power. The remaining $60\%$ of power is cascaded into network fields. Hence, network power also exhibits cycle dependence but is weaker than ARs. The cascade ratio remains constant between various solar cycle phases. This further implies that the diffusion process from ARs to network regions is similar through multiple scales and independent of solar cycles.

Based on Equation (\ref{eq3}), the variation of power-law indices could be explained. During the active phase, the AR power cascades to the network in a fixed ratio. Meanwhile, the approximate constant power is also injected from ERs or internetwork fields to magnetic networks. As a result, the power-law index deviates from -1.5, the value proposed by \cite{Kraichnan1965}. The influence of power injection from small scale structures increases as the AR power weakens, causing the power-law index to increase and deviate further from -1.5. When ARs rarely emerge on the surface, ER and internetwork fields dominate the network, resulting in positive indices.

Figure \ref{fig:fig8} (c) shows the variation of ratios between network power and total power. This is another representation of the weak cycle dependence of the network. The ratio varies slightly, with the largest value being $22.1\%$ and the smallest being $19.3\%$. Taking account of the small amplitude of ratio variation, we can estimate that the ratio is approximately $20\%$ during any phase of solar cycles.

\subsubsection{Similar Magnetic Power Spectra for Magnetograms without ARs}\label{subsubsec:minimum}

\begin{figure}[ht]
    \centering
    \includegraphics[width=1.0\linewidth]{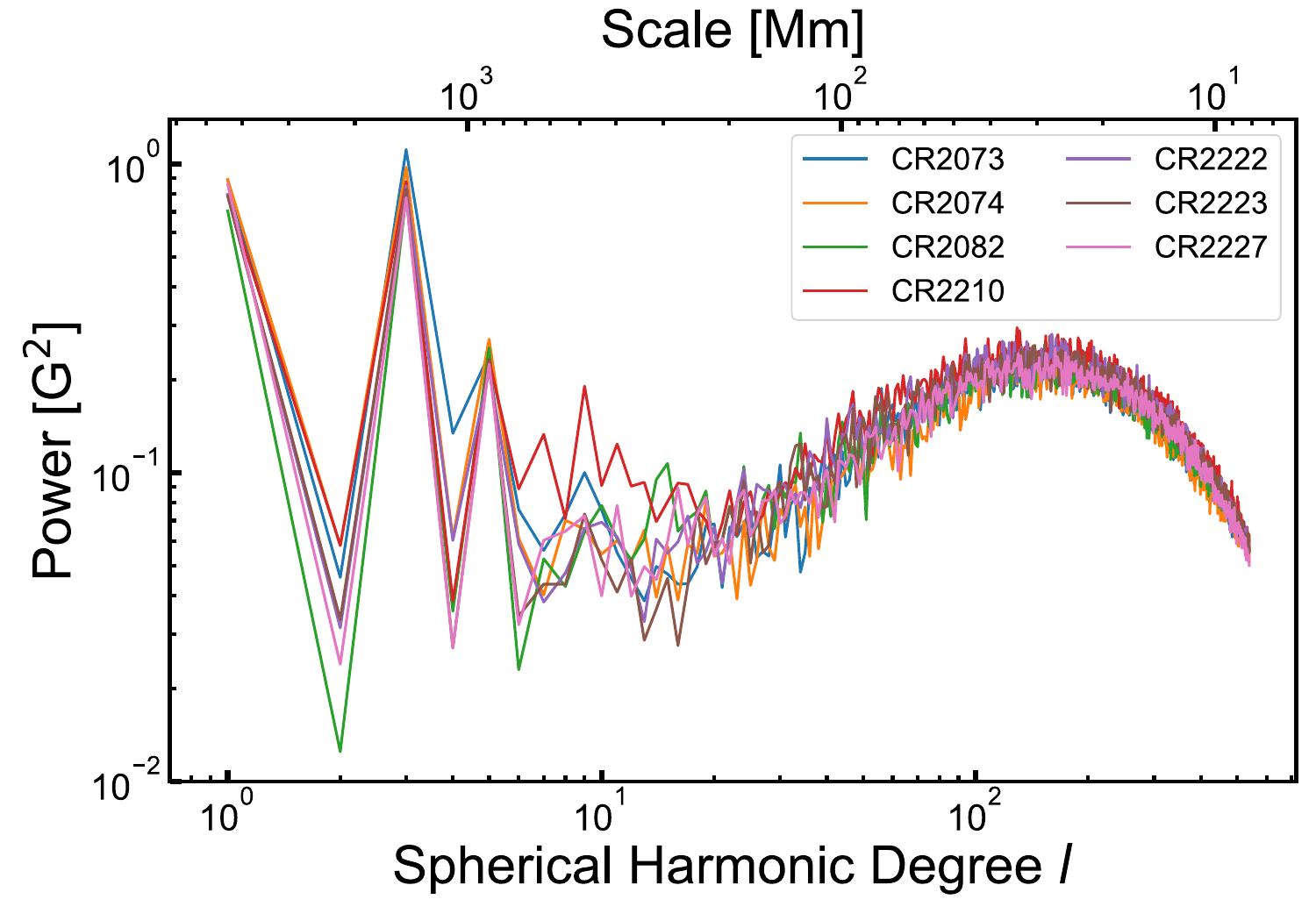}
    \caption{Magnetic power spectra of the 7 CRs of which synoptic magnetograms without any ARs: CRs 2073 (blue), 2074 (yellow), 2082 (green), 2210 (red), 2222 (purple), 2223 (brown), and 2227 (pink).}
    \label{fig:fig10}
\end{figure}

Figure \ref{fig:fig10} displays the seven magnetic power spectra from various solar minimums without ARs. Their average power spectrum is presented in Figure \ref{fig:fig5}. All power spectra for $l>40$ exhibit similar profiles and power, and they roughly overlap. Their power-law indices are around $0.72\pm0.08$, slightly higher than that in the quiet-Sun spectrum proposed by \cite{Abramenko2001}. Similarly, \cite{Foukal2001} analyse Ca \uppercase\expandafter{\romannumeral2} photographic plates and suggest that the network area does not vary significantly among the nine solar minima in the last century. In kinetic power spectra, the peaks corresponding to supergranulation are also nearly constant \citep{Williams2011b}. We can speculate that the network has similar properties when there are no ARs on the solar surface.

Considering that during the grand minima, such as the Maunder minimum \citep{Spoerer1890,Usoskin2023}, ARs rarely appear on the surface, the magnetic power spectra of the 7 CRs shown in Figure \ref{fig:fig10} could be applied to the grand minima. Thus, the magnetic power spectra for spotless days provide an effective way to perceive the magnetic field, and consequently, the solar irradiance \citep{Dasi-Espuig2016}, etc. during the grand minima.

\section{Conclusion and Discussion} \label{sec:conclusion}

This paper presents a new approach to measuring both the AR and magnetic network sizes, as well as investigating their cycle-dependent properties and relationships. The results show that the size of ARs ranges from 78 Mm to 365 Mm and their corresponding areas follow a log-normal distribution. The identified network sizes range from 26 Mm to 41 Mm, which are close to the supergranulation sizes determined from kinetic power spectra by \cite{Williams2014,Hathaway2015}. But \cite{Derosa2004} and \cite{Meunier2007c} obtain different values for the supergranulation size using the local correlation tracking method, which shows a pronounced dependence on the smoothing procedure \citep{Rincon2018}. Our results indicate that the typical network size has no cycle dependence. During the active phases of cycles 23 and 24, the network is identified in 87 CRs, suggesting that it is a significant feature not only in the quiet sun. 

We also study the cycle dependence of the network power ($P_{NW}$) and its relation with AR power ($P_{AR}$). We find that the power-law index between AR sizes and network sizes displays an anti-correlation with solar cycles. The ratio between network power and total power is approximately 20$\%$ regardless of the cycle phase. The network power shows a weaker cycle dependence than the AR power, their relationship is described by $P_{NW}$ $\approx$ 0.6* $P_{AR}$ + 40. Based on this relationship, we propose a possible explanation for the variation of the power-law index. The two terms on the right side of the equation might correspond to two different sources of the typical magnetic network and decaying ARs could be the primary source during the solar maximum.

In addition, we find that in the absence of ARs on the solar surface, the power spectra are time-independent and exhibit similarity in shape and power. We propose that the magnetic power spectra might exhibit comparable features during the grand minima such as the Maunder minimum when ARs rarely appear on the solar surface. The magnetic power spectra of the synoptic maps without ARs, as presented in Figure \ref{fig:fig10}, might be applicable to the grand minima, particularly for the spectra with $l>\sim20$. Hence, these power spectra provide a new way to estimate properties of magnetic fields during the grand minima.

The typical network sizes identified are cycle-independent, but the cycle dependence of supergranulation size is controversial. \cite{Meunier2007} propose that strong magnetic network fields are typically associated with relatively larger supergranulation. Since supergranulation sizes and magnetic network sizes could have certain differences as reported by \cite{Hagenaar1997,Rincon2018}, the supergranulation sizes could increase slightly with the solar cycle, while the typical network sizes remain cycle independent.

The network field is one of the important contributions to the variation of total solar irradiance (TSI), which affects Earth's climate \citep{Solanki2013}. The cycle dependence of network power can help to understand how the network affects TSI. While the network power exhibits a correlation with solar cycles, its ratio to total power is approximately $20\%$ for any cycle phase. We believe that this correlation also applies to other solar cycles, helping to calibrate different TSI composites and providing a new constraint for the historical TSI reconstruction \citep{Dasi-Espuig2014,Dasi-Espuig2016}. Additionally, the power spectra without ARs can be used to estimate the variation of TSI during grand minima.

This paper gives typical magnetic power spectra during solar maximum and minimum obtained directly from magnetograms. Comparing these spectra with kinetic power spectra like \cite{Hathaway2000} and \cite{Gizon2012} could reveal the relation between kinetic and magnetic energy across various scales, from global to network scales \citep{Abramenko2011}. These help for high-resolution simulation of the global magnetic field at the solar surface \citep{Hotta2022}. Additionally, the spectral features of ARs and the network, as well as their cycle dependence, can help to reconstruct high-resolution magnetograms from low-resolution ones. This will be the future work.\\ \hspace*{\fill}

We are deeply indebted to the anonymous referee for the careful and invaluable comments, which helped us to improve our manuscript. The research is supported by the National Natural Science Foundation of China No. 12350004, No. 12173005, and National Key R\&D Program of China No. 2022YFF0503800. We would like to express our gratitude to the teams responsible for the development of Python toolkits such as `pyshtools' and `scipy'. The SDO/HMI data are courtesy of NASA and the SDO/HMI team. SOHO is a project of international cooperation between ESA and NASA. The sunspot number data are provided by WDC-SILSO, Royal Observatory of Belgium, Brussels.

\begin{appendix}\label{sec:appedix}

\begin{figure*}[ht]
    \centering
    \includegraphics[width=1.0\linewidth]{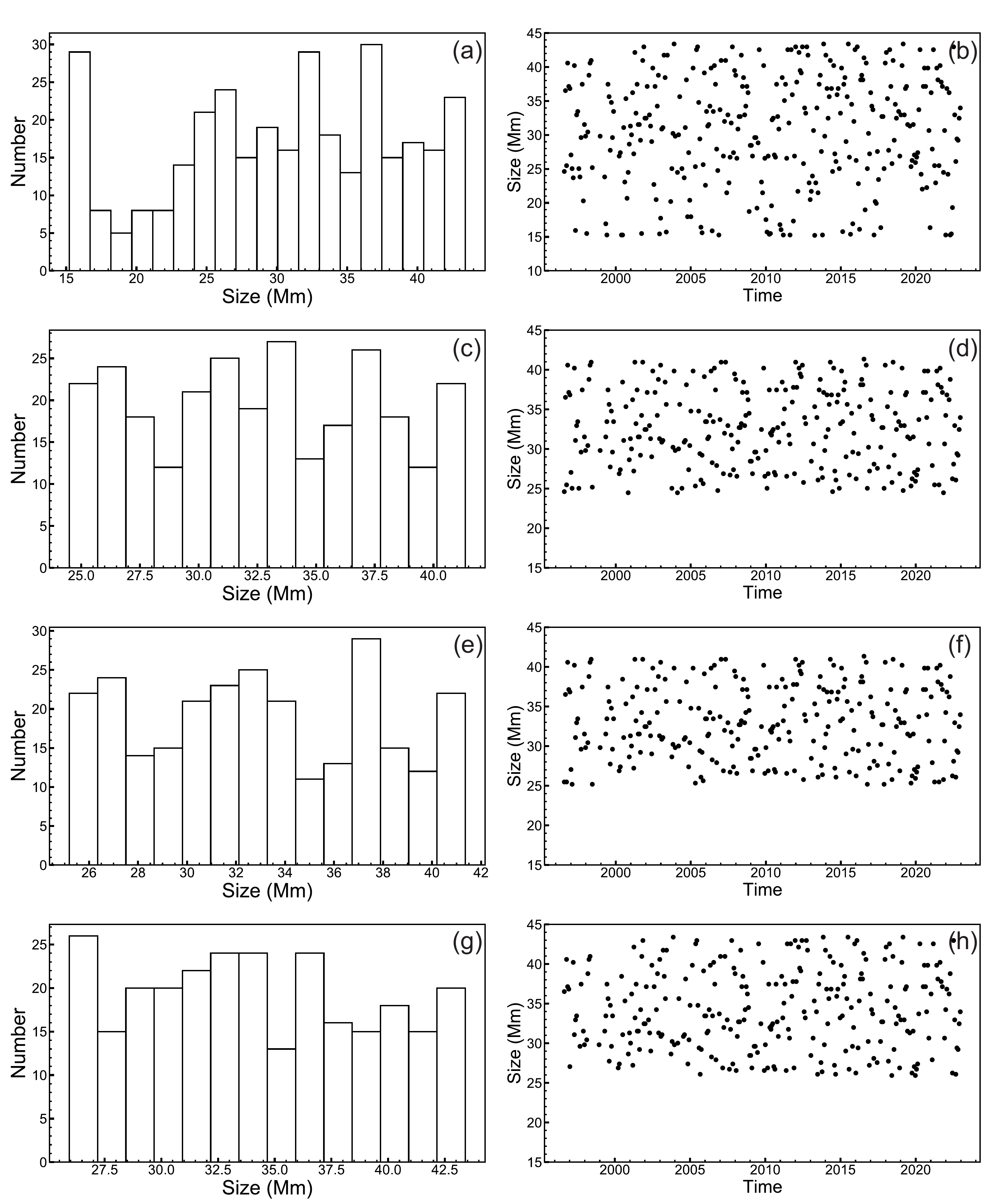}
    \caption{Identified network sizes (right) and their histograms (left) using different identification ranges of $l$. From top to bottom, the ranges of $l$ are 100-300, 105-180, 105-175, and 100-170, respectively.}
    \label{fig:fig11}
\end{figure*}

Figure \ref{fig:fig11} shows the identified network sizes and their histograms using different identification ranges of $l$ comparing with the range used in Section \ref{subsec:method}. The top panel corresponds to a broad identification range: $l=100\sim300$ (14.6 $\sim$ 43.8 Mm), and we get the network sizes from 15.2 to 43.4 Mm. Panel (a) shows that the distribution of results exceeding 25 Mm is also nearly uniform, whereas the portion below 25 Mm is not. In panels (c)-(f), using the various upper limits of $l$=180, 175, the lower limits of identified network sizes extend to 24.5 Mm and 25.2 Mm, respectively. In panels (g) and (h), we change the lower limit of $l$ from 105 to 100 and get the upper limit of the identified network size of 43.4 Mm. In all cases, the identified network sizes are not dependent on the solar cycle.

\end{appendix}

\bibliography{power_spectra}{}
\bibliographystyle{aasjournal}



\end{document}